\documentclass[vecphys]{svmult}

\usepackage{makeidx}         
\usepackage{graphicx}        
\usepackage{multicol}        
\usepackage[bottom]{footmisc}

\makeindex             

\begin{document}

\title*{Gas flows and bars in galaxies}
\author{F. Combes}
\institute{LERMA, Observatoire de Paris, 61 Av. de l'Observatoire,
F-75014, Paris, France \texttt{francoise.combes@obspm.fr}}
%
%
\maketitle

\begin{abstract}
Dynamical evolution of spiral galaxies is strongly dependent
on non-axisymmetric patterns that develop from gravitational
instabilities, either spontaneously or externally triggered.
  Some evolutionary sequences are described
through which a galaxy could possibly concentrate mass and
build bulges, how external gas accretion from cosmic
filaments could be funneled to the galaxy disks, and intermittently
driven to the galaxy center, to form nuclear starbursts and fuel
an active nucleus. The frequency of both bars and lopsidedness
can be used to constrain the gas accretion rate. 
\end{abstract}

\section{Evidence of gas flow, central star formation and pseudo-bulge formation}

\subsection{Molecular gas in barred galaxies}

Sakamoto et al (1999) from a sample of 20 nearby galaxies were the first to notice that
barred galaxies had a larger
gas concentration in the central kiloparsec than unbarred galaxies, confirming
that bar instabilities were able to drive gas inflow towards the center. 
Sheth et al (2005) from a sample of about 45 galaxies from BIMA-SONG confirm
the larger CO concentration for barred galaxies, and find that the nuclear molecular
gas is 4 times more abundant in early-types than late-types, suggesting the gas inflow
are larger, possibly due to the existence of inner Lindblad resonances (ILR), expected
when the mass is concentrated due to bulges.
Jogee et al (2005) further note that barred galaxies have a large 
circumnuclear surface densities in molecules, and this triggers the nuclear
starbursts currently observed there. 
The observation of non-starbursting gas-rich barred galaxies allows
to distinguish several phases of evolution:
the first phase reveals gas inflow, non
circular motions (since the gas is partly in the bar),
and the gas is accumulating at the outer ILR, there is no starburst. 
In the end-phase, the gas has been driven down
to the inner ILR, runs in nearly circular orbits, and
the galaxy experiences a nuclear starburst of 3-11 M$_\odot$/yr.
Molecular gas surface densities reach up to 500-3500 M$_\odot$/pc$^2$, or
10-30\% of dynamical mass in the center.

\subsection{Star formation in the center of spiral galaxies}

\begin{figure}
\centering
\includegraphics[width=0.7\textwidth]{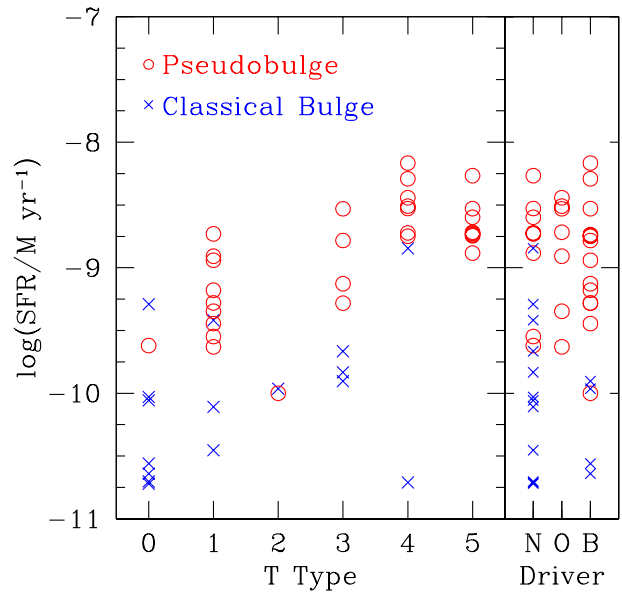}
\caption{Specific SFR of the central 1.5 kpc, for a sample of 50 galaxies, from Fisher (2006), as a function 
of morphological type (left), and as a function of secular driving mechanism (B=bar, O=oval, and 
N=neither bar no oval). Pseudo-bulge galaxies (red circles) have clearly larger central specific SFR
than classical bulge ones (crosses).}
\label{fisher}  
\end{figure}

Recent work on star formation in the central kpc of galaxies has
shown that galaxies with pseudo-bulges are more frequently the host
of starbursts. Pseudo-bulges are components intermediate between spheroids
and disks, in the sense of light distribution (Sersic index $n$ between 1 and 2)
of flattening and kinematics (more rotation than classical bulge,
Kormendy \& Kennicutt 2004). Fisher (2006) has recently made a star-formation
study on a sample of 50 galaxies observed by Spitzer. Pseudo-bulges were identified 
 on HST images from the presence of disky structure in the center, i.e. nuclear disks
nuclear bars, and high ellipticity. The star formation was traced from
the PAH emission at 8 micron wavelength. As shown in Fig \ref{fisher},
star formation inside 1.5 kpc is clearly larger in pseudo-bulge galaxies,
and among them, central starbursts are preferentially
in barred galaxies, then in galaxies with oval perturbations.
Nuclear star formation in the form of H$\alpha$ nuclear rings are also 
preferentially found in barred galaxies (Knapen 2005, and this meeting).

Peeples \& Martini (2006) study the
relation between bar strength (in terms of Q$_b$) and circumnuclear dust. Dust is distributed
preferentially in nuclear rings for barred galaxies.
SB(r) galaxies are less strongly barred and have less dust structure than SB(s) galaxies. 
This can be understood as an evolutionary process, since rings are formed in a second phase
in barred galaxies, where first a spiral structure forms.

\subsection{Bar length and type, $\Omega_b$ and dynamical friction}

\begin{figure}
\centering
\includegraphics[width=0.8\textwidth]{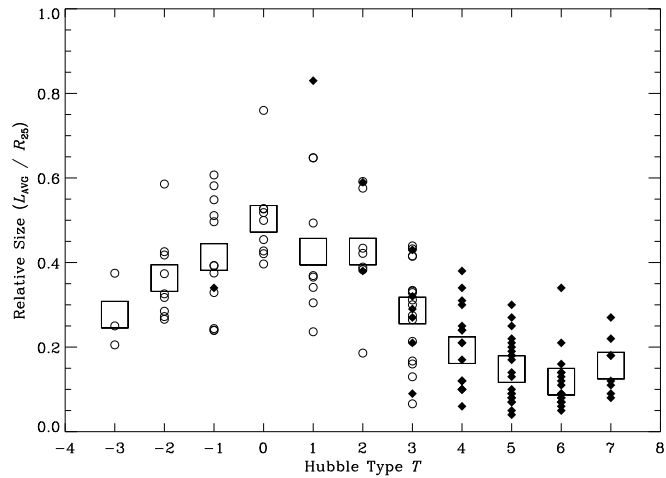}
\caption{Sizes of bars (deprojected) relative to their disc radius R$_{25}$,
as a funciton of morphological type, from Erwin (2005).
The various symbols correspond to various samples.
Large boxes are mean values for each type.}
\label{erwin}  
\end{figure}

In a recent study of about 140 galaxies, Erwin (2005) confirms that bars are
longer in early-types than in late-type galaxies, both when normalized with 
the optical radius R$_{25}$, or with the exponential scale-length $h$ 
(see Figure \ref{erwin}). The size of bars does not depend on their strength.

These observational results can be confronted to predictions from
numerical models. The length of bars is tightly related to their pattern speed,
since stellar orbits support the bar until its corotation only.
 In the presence of a large mass of dark matter inside the bar radius,
bars should slow down through dynamical friction against the dark matter particles
(Debattista \& Sellwood 2000), and bars should then be too long and slow
compared to observations.  This puts constraints on the dark matter
distribution in spiral galaxies, and favors maximal disks.
But in some cases, the transfer of angular momentum to
the dark halo can be minimized (Valenzuela \& Klypin 2003).

The length of bars is interesting under the point of view of secular evolution,
since it is predicted to vary along the successive evolution phases, and is then
 a tracer of the past history of the galaxy.  Bars begin their life with a high
pattern speed and a short length (as in late-types) and then slow down and
increase their length, as in early-types. 
This simple scenario is slightly modified when bar destruction and 
reformation is taken into account: 
with gas accretion, new bars are generally shorter. They appear late-type
just after gas accretion (Bournaud \& Combes 2002). Tidal 
interactions on the contrary can produce longer bars (Holley-Bockelmann et al 2005),
depending on mass and distance of the satellites.

In edge-on galaxies, the length of bars can be traced by
the size of the peanut-shape. The maximum height of the peanut
occurs at the z-resonance with the bar, a function of its pattern speed.
When the bar slows down through angular momentum transfer with the halo,
the radius of the resonance shifts outwards. Several peanut buckling can occur
successively (Martinez-Valpuesta et al 2006).

\section{Bar destruction and reformation}

Along the secular evolution, bars are robust and long-lived features
only in pure stellar disks. When gaseous disks are taken into account
in numerical simulations, bars are observed to weaken and disappear
(e.g. Friedli et al 1994, Bournaud \& Combes 2002).
This was long attributed to the building of a central mass concentration,
due to the gas inflow towards the center (Hasan et al 1993).  However,
when the central mass concentration is considered alone, it does
not destroy the bar (Shen \& Sellwood 2004). The details of the evolution
are now much better understood:  gas is driven inwards by the bar torques.
The gas angular momentum is taken up by the bar wave. The latter
has a negative angular momentum, being contained inside its corotation (CR). 
Providing positive angular momentum to the wave  is sufficient 
to weaken or destroy the bar, since the total gas angular momentum
from CR to center is of the same order as the bar wave momentum
(Bournaud et al 2005a).

Since it is not the central mass concentration which destroys the bar,
it is relatively easy to reform a bar, after a bar episode is completed.
Substantial gas accretion is required to replenish the disk, and to make
it cold and again unstable to bar formation (Block et al 2002). Several bar episodes
can then occur in a given galaxy, with a time scales of a few Gyr each,
provided that the galaxy mass is doubled in 10 Gyr typically, which is 
compatible with cosmological accretion rates. 

It is interesting to note that the angular momentum transfer mediated
by the bar, redistributes the mass in galaxies and produces exponential
light profiles, as observed in spiral galaxies. Recently, several exponential
components have been identified in galaxy disks, separated by 
breaks in the density profiles (Pohlen 2002). These could be the 
consequence of the past activity of several bars along the secular evolution.

Secular evolution with continuous gas accretion can also trigger 
starbursts, and mimick galaxy interactions. Gas is accreted from CR onto the central
regions only intermittently, when the bar is at its maximum strength.
Gaseous nuclear disks are then formed in a short time-scale, and starbursts 
may be triggered. Repeating starbursts have been observed in some barred galaxies
(cf Allard et al 2006).
In the mean time, external gas continuously accreted accumulates in the outer
parts, and can inflow to replenish the disk only when the bar has been weakened.
Just after gas accretion, which is in general asymmetric, galaxies look 
peculiar: lopsidedness and warping structures can not only be due to galaxy-galaxy 
interactions but also to asymmetric mass accretion.

\section{Gas accretion and $m=1$ perturbations}

The quantification of observed bar frequency from the near-infrared images 
of the OSU sample (Eskridge et al 2002) has been used to put constraints
on the gas accretion rate (Block et al  2002).  The same can be done for the
lopsided perturbations (or Fourier $m=1$).  This asymmetrical perturbation
is very frequently observed in galaxies, even in the absence of any
companion. Half of the HI profiles in sample of 1700 galaxies have been
found  asymmetric (Richter \& Sancisi 1994), and these perturbations
are even more frequent in late-type galaxies (Matthews et al 1998).
Stellar disks  also participate in those asymmetries (Zaritsky \& Rix 1997).
About 20\% of galaxies have the normalized $m=1$ Fourier
coefficient in near-infrared surface density A1 larger than 0.2, 
and in about 2/3 of galaxies,  an external mechanism is required to explain
these asymmetries, that are above the expected amplitude of
spontaneous internal perturbations.

\begin{figure}
\centering
\includegraphics[width=0.8\textwidth]{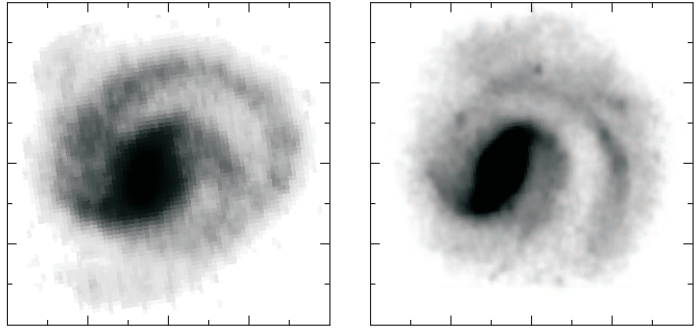}
\caption{Comparison between a lopsided isolated galaxy NGC~1637 
(right: NIR map from the OSUBGS data, after deprojection), and a numerical
simulation of asymmetrical gas accretion (left), producing a
strong lopsidedness, from Bournaud et al (2005b). The gas accretion rate is of  6~M$_{\odot}$~yr$^{-1}$, 
and for a galactic mass of $10^{11}$~M$_{\odot}$~yr$^{-1}$ correspond
to doubling the mass of the galaxy in about one Hubble time.}
\label{1637}  
\end{figure}

The influence of companions however does not appear
to be the solution:  most lopsided galaxies are isolated 
(Wilcots \& Prescott 2004), and when companions are seen,
the  A1 parameter  is not correlated with the
tidal index, computed from the mass ratio between the observed
neighbours and the target, and  inversely proportional to the cube 
of their distance. In any case, interactions cannot
explain that A1 is higher in late-type galaxies
(Bournaud et al 2005b). External gas accretion,
which is likely to be asymmetric at a given epoch, 
is able to explain the asymmetries observed. 
Fig  \ref{1637} illustrates the mechanism,
with a comparison between an observed asymmetric case and
a numerical simulation. Gas accretion can
explain the long life-time of the  perturbation,
and consequently the high frequency of lopsidedness
observed, while a minor merger for instance will produce
asymmetries on a much shorter time-scale.

\printindex
\end{document}